\documentstyle[twoside,fleqn,tk]{article}

\newcommand{\AmS}{{\protect\the\textfont2
  A\kern-.1667em\lower.5ex\hbox{M}\kern-.125emS}}

\title{{
\begin{flushright}
{\normalsize
PDK-742 \\
TUHEP-99-05 \\
\vskip -6pt
14 Dec. 1999}
\end{flushright}
{
Recent atmospheric neutrino results from Soudan 2}}
 \thanks{Presented at TAUP99, the 6th International 
Workshop on Topics in Astroparticle and Underground Physics, Sept. 6-10,
1999, College de France, Paris, France.}}

\author{T. Kafka \thanks{High Energy Physics, Tufts University,
4 Colby St., Medford, MA 02155, USA}
\vskip 0.2cm
\noindent {\it for the Soudan 2 Collaboration}
\vskip 0.1cm
\noindent (Argonne National Laboratory, University of Minnesota, Tufts
University, USA; Oxford University, Rutherford Appleton Laboratory, UK)}

\begin{document}


\begin{abstract}
An updated measurement of the atmospheric $\nu_\mu/\nu_e$ ratio-of-ratios,
$0.68\pm 0.11\pm 0.06$, has been obtained using a 4.6-kty exposure of the
Soudan-2 iron tracking calorimeter. The $L/E$ distributions have been
analyzed for effects of $\nu_\mu \rightarrow \nu_x$ oscillations, and
an allowed region in the $\Delta m^2$ vs. sin$^2 2\theta$ plane
has been determined.

\end{abstract}

\maketitle

\section{INTRODUCTION}


Soudan 2 is a fine-grained iron tracking calorimeter located 2100 mwe
underground in Soudan, Minnesota, USA. The experiment 
has been taking data since 1989 when the detector was one
 quarter of its full size; the construction was completed in 1993.
The experiment continues to take data with 90\% live time. The results
presented here were obtained using a 4.6 fiducial kiloton-year exposure.
For previous Soudan-2 results on the ratio of ratios of the atmospheric
 neutrinos,
\begin{equation}
 R = {{[(\nu_\mu+\bar\nu_\mu)/(\nu_e+\bar\nu_e)]_{Data}}
 \over{[(\nu_\mu+\bar\nu_\mu)/(\nu_e+\bar\nu_e)]_{MC}}} \label{eq:R}
\end{equation}
see Refs. \cite{First,Second}.

Detailed descriptions have been published
of the design and performance of Soudan-2 calorimeter modules
\cite{NIM}, and of the active shield system \cite{NIM_VS}.

\section{ATMOSPHERIC NEUTRINO FLAVOR RESULT}

We analyze fully contained events (all hits more than 20 cm from the nearest
detector edge) to determine $R$ using single-track and single-shower
events for which the active shield registered no in-time
two-layer coincident hits (gold events).
 This sample contains mostly quasi-elastic neutrino interactions,
with a background of photon and neutron interactions originating in
cosmic-ray muon interactions in the rock. These rock events are mostly
flagged by hits in the active shield, but some are not accompanied
by shield hits and constitute a background for our neutrino sample.
A correction for this background is determined using distributions of
event depth in the detector. For the 4.6-kty Soudan 2 exposure, we obtain
a ratio-of-ratios value of
$$ R = 0.68 \pm 0.11(stat.) \pm 0.06 (syst.)$$
Our full detector simulation of atmospheric neutrino interactions is based on
the 1989 Bartol flux for the Soudan site \cite{Bartol}.
See Ref. \cite{TAUP97} for more information on the quasi-elastic data analysis.

\section{HIGH RESOLUTION SAMPLE}

We wish to analyze neutrino $L/E$ distributions where the neutrino path length
is calculated from the event zenith angle, radius of the Earth, and the
neutrino production height.
We therefore make data selection designed to optimize our resolution in
event angle and in energy.
 For the quasi-elastic events (tracks, showers) we
require the charged lepton momentum to exceed 150 MeV/c if a recoil proton is
measured in the event, otherwise we require event visible energy to exceed
600 MeV. For multiprongs, we require the visible energy to exceed 700 MeV,
the vector sum of visible momenta to exceed 450 MeV, and the
charged-lepton momentum to exceed 250 MeV.
 The event counts in the resulting
event samples are given in Table 1. 

\begin{table}[hbt]
\label{tab:HiRes}
\caption{Soudan-2 HiRes event samples for the atmospheric neutrino data
(before and after the background subtraction) and Monte Carlo.}
\setlength{\tabcolsep}{0.7pc}
\begin{tabular}{@{}l@{\extracolsep{\fill}}ccc}
\hline
& & $\nu_\mu$ & $\nu_e$ \\
\hline
 Data, before & & 99.0$\pm$10.0 & 121.0$\pm$11.0 \\
 Data, after  & & 92.3$\pm$10.1 & 114.6$\pm$11.1 \\
 Monte Carlo  & & 133.6$\pm$4.4 & 114.6$\pm$4.1 (norm.) \\
\hline
\end{tabular}
\end{table}

We calculate our `resolution' as the difference between the Monte Carlo truth
and the reconstructed final-state characteristics, thereby taking
into account both the event kinematics and the instrumental resolution.
 The resulting $\Delta E/E$ is 20\%, and the average angular resolution is
 33$^\circ$ for $\nu_\mu$, and 21$^\circ$ for $\nu_e$ charged-current events.
 The corresponding resolution in log$(L/E)$ is 0.49 and 0.43, respectively.

In this `high-resolution' (HiRes) sample, the rock background is small, 5--7\%,
and the neutrino flavor is determined correctly 92\% of the time. The
Monte Carlo sample represents 25 kty of exposure.

\section{$L/E$ RESULTS}

\begin{figure}[hbt]
\vspace {210.0pt}
\includegraphics{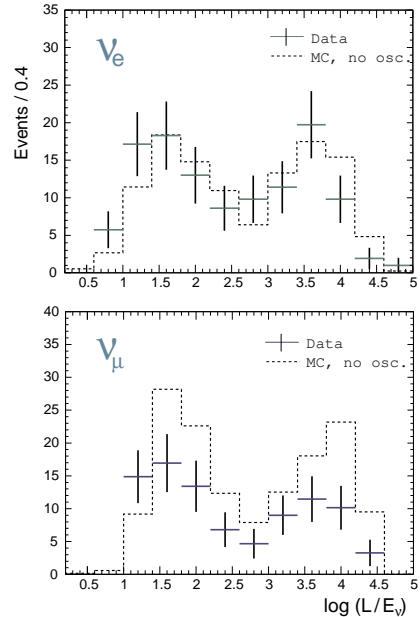}
\caption{$L/E$ distributions for Soudan-2 $\nu_e$- and $\nu_\mu$-flavor
 HiRes samples (crosses) together with the
no-oscillation atmospheric neutrino Monte Carlo (dashed histograms).}
\label{fig:lovre_noosc}
\end{figure}

In Fig. 1 we show the $L/E$ distributions for the HiRes sample.
Fig. 1a shows that the $\nu_e$-flavor data (crosses) follow the shape of
the  $\nu_e$-flavor no-oscillation Monte Carlo distribution (dashed histogram).
(The MC histogram is normalized to the $\nu_e$ data.) 
In the $\nu_\mu$-flavor sample, shown in Fig. 1b, the depletion of the data
vis-a-vis the no-oscillation Monte Carlo is obvious (with the same
MC normalization as for the $\nu_e$ sample).

\begin{figure}[hbt]
\vspace {200.0pt}
\includegraphics{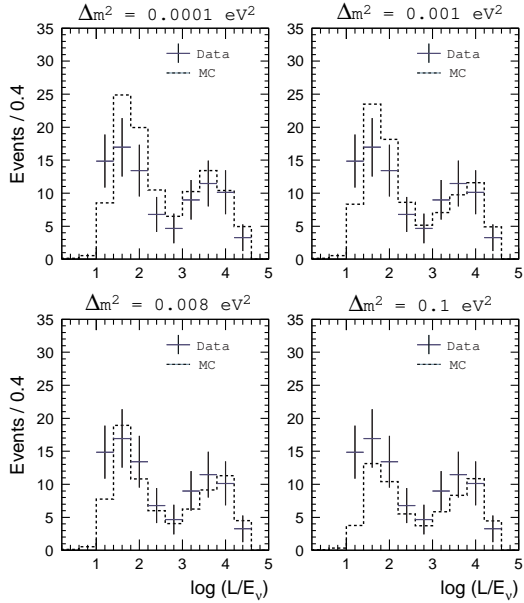}
\caption{$L/E$ distribution for Soudan-2 $\nu_\mu$-flavor
HiRes sample (crosses) together with the
atmospheric neutrino Monte Carlo with oscillations for sin$^2 2\theta = 1$
and for four values of $\Delta m^2$ (dashed histograms).}
\label{fig:lovre_osc}
\end{figure}

To convert results of our atmospheric neutrino simulation generated under the
no-oscillation hypothesis into simulated neutrino oscillation data, we apply to 
every MC event an $L/E$-dependent weight representing the probability of
no oscillation for a given $\Delta m^2$ and sin$^2 2\theta$ valid for
two-flavor neutrino oscillations. Fig. 2 displays the $\nu_\mu$-flavor
$L/E$ distribution (crosses) together with the oscillation MC for
sin$^2 2\theta$ = 1 and for four values of $\Delta m^2$. We see that the
simulation exceeds the data for
$\Delta m^2$ = 0.0001 eV$^2$ and 0.001 eV$^2$ for most of the downgoing
neutrinos, while the simulation is smaller than the data for
$\Delta m^2$ = 0.01 eV$^2$. The best agreement between data and the
simulation is obtained for  $\Delta m^2$ = 0.008 eV$^2$.

To determine the neutrino oscillation parameters  $\Delta m^2$ and
sin$^2 2\theta$ from our data, we fit the atmospheric neutrino
Monte Carlo distribution including the effects of neutrino oscillations
to the $L/E$ distribution for our $\nu_\mu$ data corrected for the rock
background, by minimizing $\chi^2$. At the same time we assume that no neutrino
oscillations occur in the $\nu_e$ flux, and include only the total $\nu_e$
event counts in our  $\chi^2$.
In addition to  $\Delta m^2$ and sin$^2 2\theta$, the
neutrino flux normalization factor, $f_\nu$, is the third free parameter
in the fit. Errors in both the data and the Monte Carlo are taken into
account when calculating $\chi^2$.

\begin{figure}[hbt]
\vspace {180.0pt}
\includegraphics{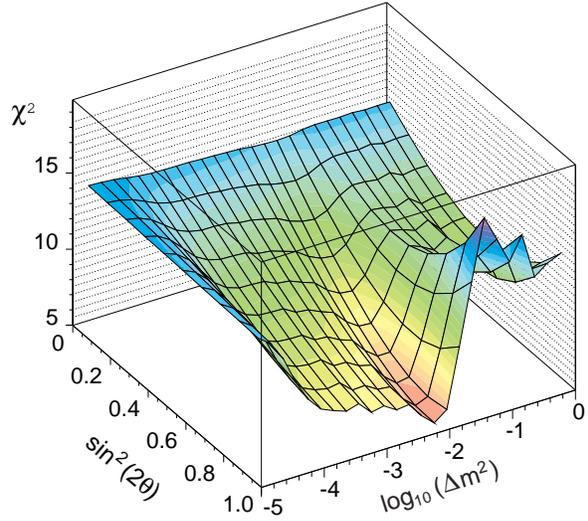}
\caption{The $\chi^2$ as a function of $\Delta m^2$ and sin$^2 2\theta$.}
\label{fig:chisq}
\end{figure}

Fig. 3 shows our $\chi^2$ surface as a function of  $\Delta m^2$
and sin$^2 2\theta$. A distinct minimum is observed at 
$\Delta m^2$ = 0.008 eV$^2$ and sin$^2 2\theta$ = 0.95 (with $f_\nu$ = 0.82).
The corresponding confidence-level contours are shown in Fig. 4 for
68\% CL and 90\% CL. The Soudan-2 allowed region is seen to overlap both
the Kamiokande and SuperKamiokande allowed regions \cite{Nakahata}.

\begin{figure}[hb]
\vspace {130.0pt}
\includegraphics{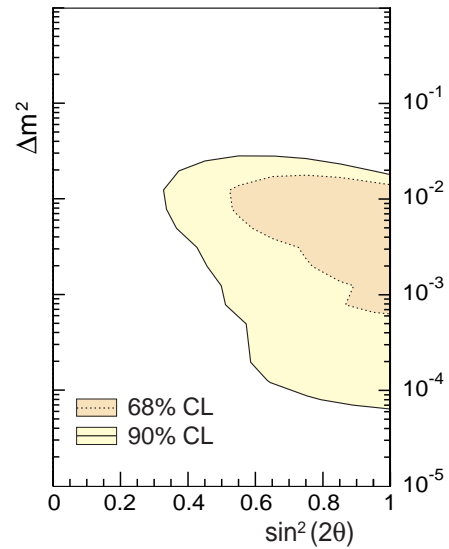}
\caption{Soudan-2 allowed regions for CL of 68\% and 90\%.}
\label{fig:CL}
\end{figure}

Our plans for the near future are to incorporate more recent atmospheric
$\nu$ fluxes in the Monte Carlo, and to include partially contained events
in the analysis. And we certainly plan on taking more data, even through the
period of dynamite blasting of the MINOS cavern in the Soudan mine.
We expect to reach 5.0 fid. kty by the Millenium.

\end{document}